\documentclass[a4paper,11pt]{article}
\pdfoutput=1 

\usepackage{jinstpub} 
\usepackage{lineno}

\usepackage{graphicx}
\usepackage{color, colortbl}
\usepackage{hyperref}
\usepackage{float}
\usepackage{subcaption}

\title{Real-Time Charged Track Reconstruction for CLAS12}

\author[a,1]{Gagik Gavalian}
\affiliation[a]{Jefferson Lab, Newport News, VA, USA}

\abstract{
This paper presents the results of charged particle track reconstruction in CLAS12 using artificial intelligence. In our approach, we 
use machine learning algorithms to reconstruct tracks, including their momentum and direction, with high accuracy from raw hits of the CLAS12
drift chambers. The reconstruction is performed in real-time, with the rate of data acquisition, and allows for the identification of event topologies in real-time.
This approach revolutionizes the Nuclear Physics experiments' data processing, allowing us to identify and categorize the experimental data on the fly, and will lead to a significant reduction in experiment data processing. It can also be used in streaming readout applications leading to more efficient data acquisition and post-processing.
}

\collaboration[a]{On behalf of CLAS12 collaboration}

\proceeding{AI4EIC2023 \\
   November 28-December 1\\
  Catholic University of America, Washington D.C.}
  
\begin{document}
\maketitle

\section{Introduction}
\indent
Artificial Intelligence (AI) is revolutionizing the way computations are handled in nuclear physics experiments, marking a significant shift in the field. Traditionally, nuclear physics has relied on complex mathematical models and extensive computational resources to simulate and analyze experiments. These processes are often time-consuming and computationally intensive, making it challenging to quickly process large volumes of data or accurately predict outcomes in unexplored scenarios.

With the advent of AI, especially machine learning (ML) and neural networks, this paradigm is changing. AI algorithms can analyze vast datasets more efficiently than traditional methods, identifying patterns and correlations that human researchers might miss. For instance, in particle detection and identification, AI can quickly sift through millions of particle collision events to find the few that are of interest, which would be impractical with conventional computational approaches.

Furthermore, AI is enabling more accurate modeling and simulation in nuclear physics. Neural networks, trained on historical experimental data, can predict the outcomes of nuclear reactions or the behavior of subatomic particles with a high degree of accuracy. This capability is particularly useful in scenarios where experiments are either too dangerous, expensive, or technically challenging to perform.

AI is also contributing to the design of experiments and optimization of detectors. By simulating different configurations, AI can suggest the most effective experimental setups, saving time and resources. Additionally, AI is being used in the calibration of instruments and data analysis, making the process faster and more reliable.

In conclusion, AI is not just supplementing traditional computational methods in nuclear physics but is transforming the field by providing faster, more accurate, and more efficient ways to handle complex experiments. This evolution is paving the way for discoveries and advancements in nuclear physics, highlighting the growing importance of interdisciplinary approaches in scientific research.

\section{CLAS12 Detector}

The CLAS12 (CEBAF Large Acceptance Spectrometer for 12 GeV) detector is a state-of-the-art experimental apparatus used in nuclear physics research. It is located at the Thomas Jefferson National Accelerator Facility (Jefferson Lab) in Newport News, Virginia. The detector is part of an upgrade to the Continuous Electron Beam Accelerator Facility (CEBAF), which increased the maximum energy of the electron beam from 6 GeV to 12 GeV. This upgrade allows for a more in-depth exploration of the structure and properties of nucleons (protons and neutrons) and the nature of the strong force that binds them together in the atomic nucleus.

\begin{figure}[h!]
\centering
\centerline{\includegraphics[width=0.7\columnwidth]{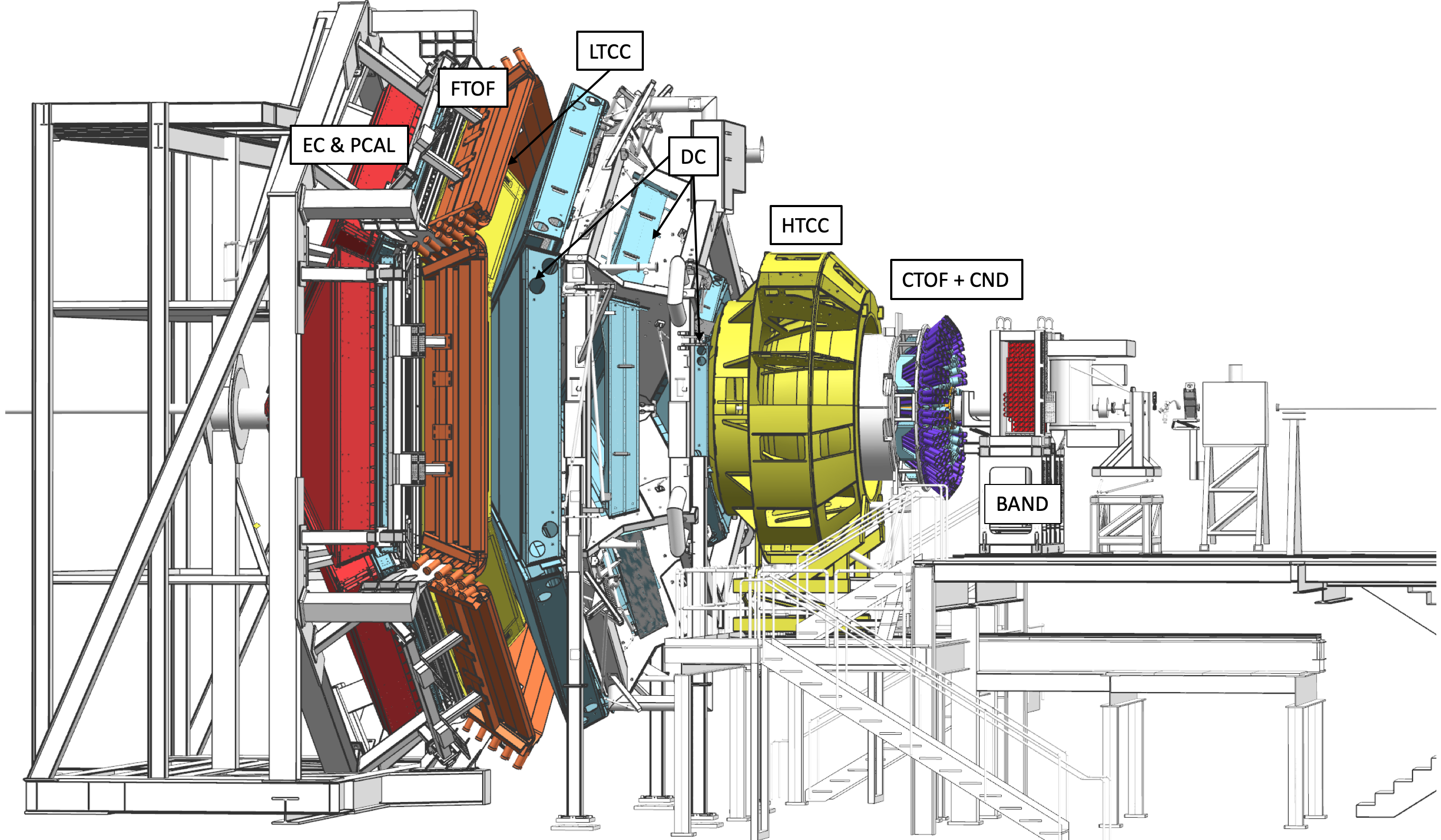}}
\caption{The CLAS12 detector in the Hall~B beamline. The electron beam enters from the right and impinges on
  the production target located in the center of the solenoid magnet shown at the right (upstream) end of CLAS12,
  where other detector components are also visible. Scattered electrons and forward-going particles are detected
  in the Forward Detector (FD), consisting of the High Threshold Cherenkov Counter (HTCC) (yellow), 
  followed by the torus magnet (gray), the drift chamber tracking system (light blue),
  and time-of-flight scintillation counters (brown), and electromagnetic calorimeters (red). } 
\label{fig:CLAS12}
\end{figure}

Key features and capabilities of the CLAS12 detector include:

\begin{itemize}

\item{\bf Large Acceptance: } As its name suggests, CLAS12 has a large angular and momentum acceptance. This feature is crucial for detecting particles over a wide range of angles and energies, allowing comprehensive analysis of nuclear reactions.
\item{\bf Electron Beam Experiments:} CLAS12 is designed to investigate the interactions of high-energy electrons with nucleons and nuclei. By scattering electrons off target materials, scientists can probe the internal structure of nucleons and the dynamics of the strong force.
\item{\bf High Luminosity:} The detector operates at high luminosities, enabling it to collect a vast amount of data from electron scattering experiments. This high data rate is essential for studying rare processes and achieving statistically significant results.
\item{\bf Sophisticated Detection Systems:} CLAS12 consists of various subsystems designed to detect different types of particles and measure their properties. These include drift chambers for tracking charged particles, time-of-flight counters for particle identification, calorimeters for measuring energy, and Cherenkov detectors for identifying electrons.
\item{\bf Versatility:} The detector is versatile and can be used for a wide range of experiments, from studying the quark-gluon structure of nucleons to investigating the properties of nuclei under extreme conditions.
\item{\bf Data Analysis and Simulation:} Advanced software and computational tools are used to analyze the data collected by CLAS12. These tools include simulation packages that model the detector's response and data analysis frameworks for extracting physical quantities from the experimental data.
\end{itemize}
In summary, CLAS12 is a critical tool in modern nuclear physics, enabling researchers to delve deeper into the quantum world of nucleons and nuclei. Its advanced technology and capabilities contribute significantly to our understanding of fundamental physics, particularly in the realm of quantum chromodynamics (QCD), the theory describing the strong interaction.

\section{Drift Chamber Particle Tracking}

\begin{figure}[h!]
\centering
\includegraphics[width=0.32\columnwidth]{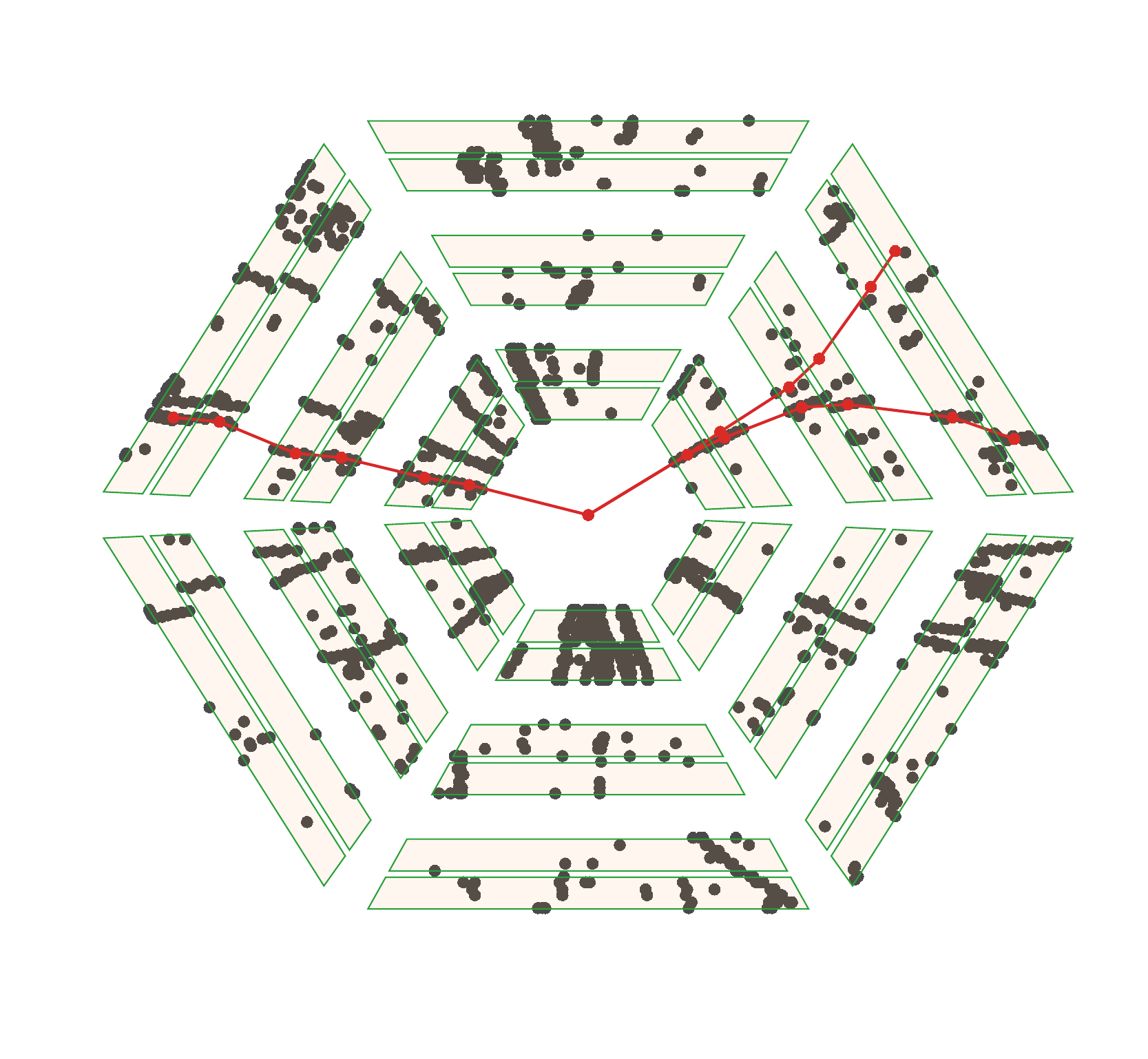}
\includegraphics[width=0.32\columnwidth]{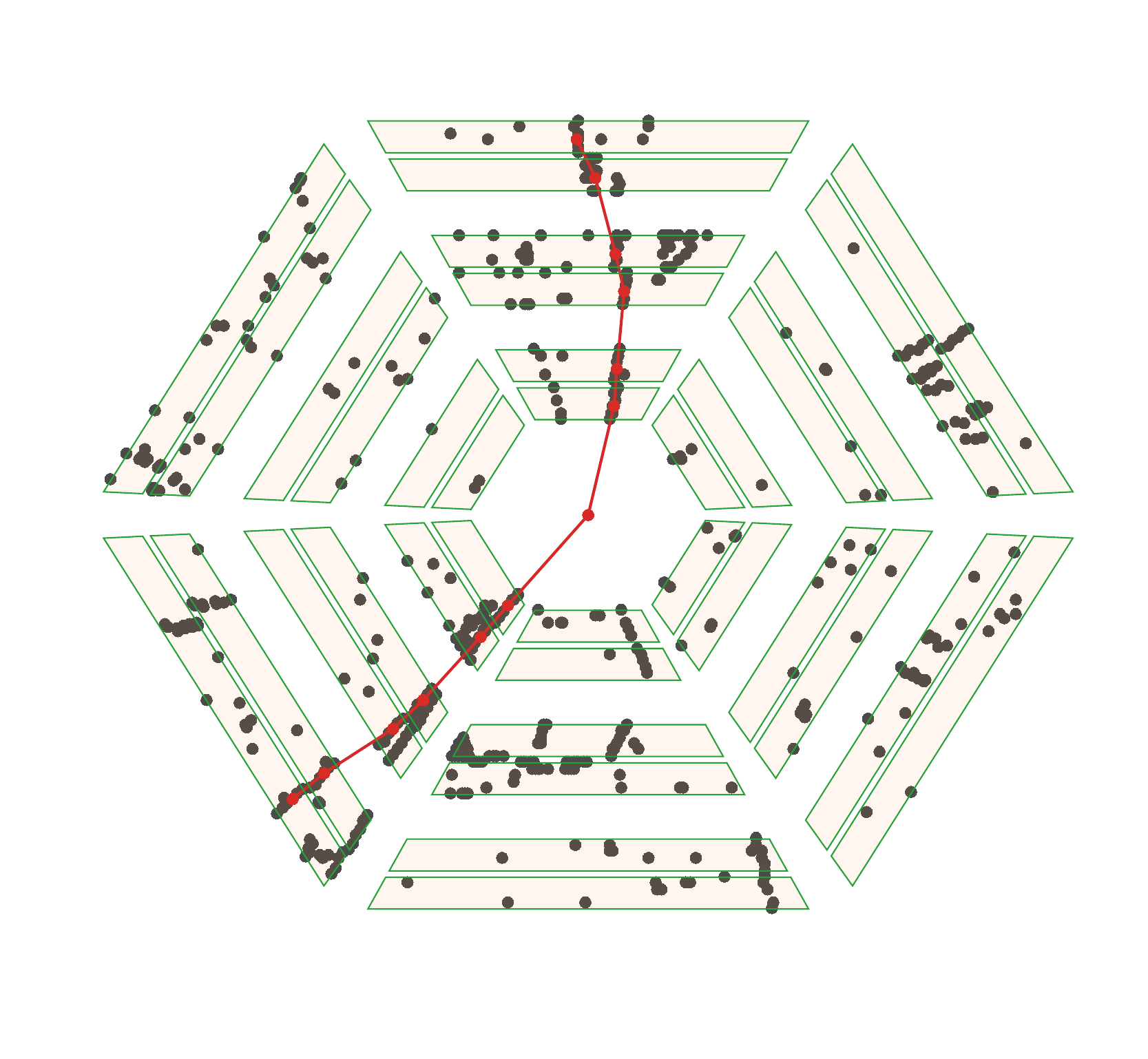}
\includegraphics[width=0.32\columnwidth]{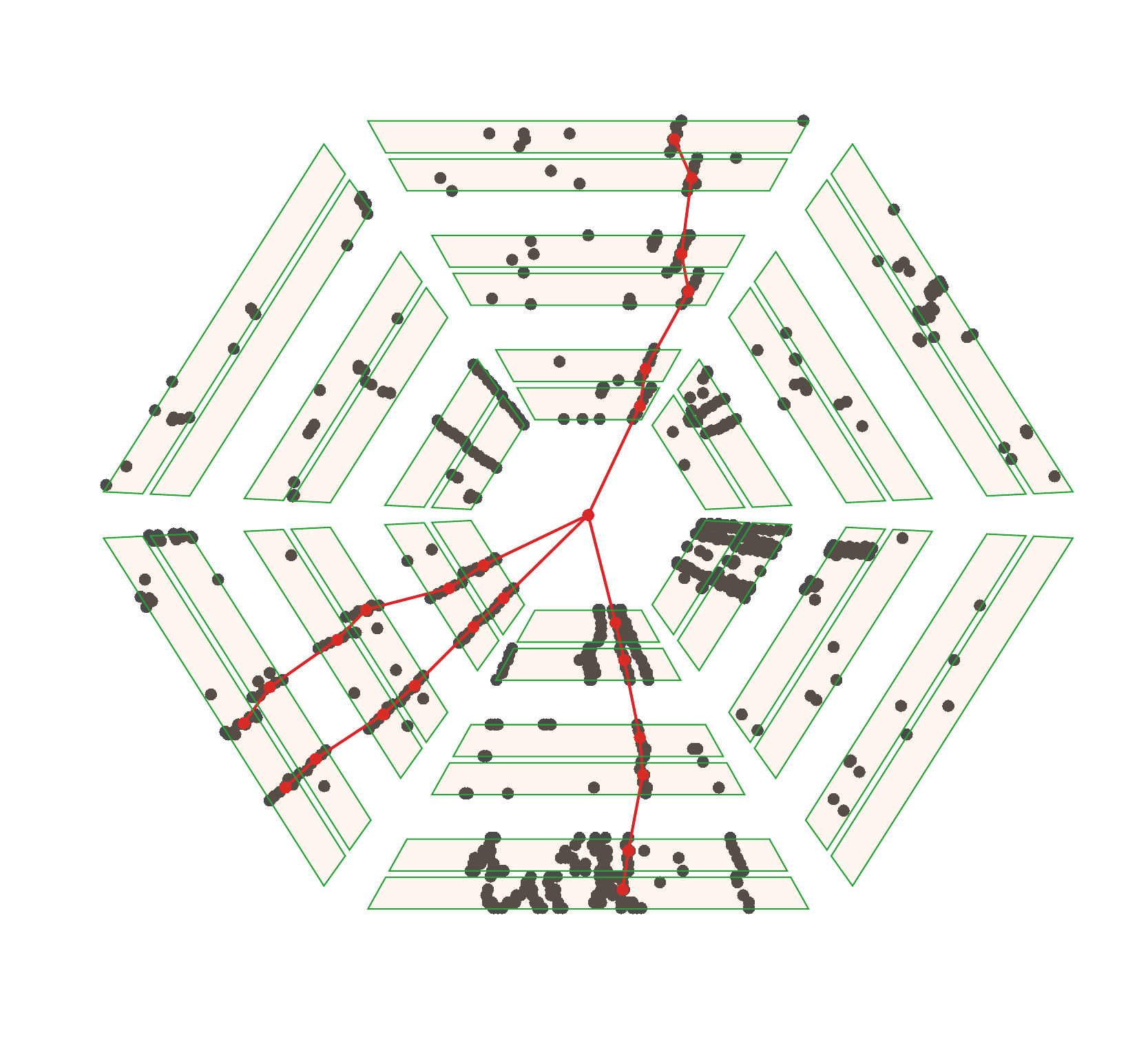}
\caption{Example views of the six sectors of the Drift Chambers of CLAS12. The points show wire hits for each of the layers in the drift chambers and the lines indicate reconstructed tracks by the conventional CLAS12 tracking algorithm. } 
\label{fig:drift_chambers}
\end{figure}

The CLAS12~\cite{Burkert:2020akg} forward detector is built around a six-coil toroidal magnet 
which divides the active detection area into six azimuthal regions, called ``sectors''. Each sector is 
equipped with three regions of drift chambers~\cite{Mestayer:2020saf} designed to detect charged 
particles produced by the interaction of an electron beam with a target. Each region consists of two 
chambers (called super-layers), each of them having six layers of wires. Each layer  in a super-layer 
contains 112 signal wires, making a super-layer a $6\times112$ cell matrix. 
The schematic view of all sectors and super-layers is shown in Figure~\ref{fig:drift_chambers}.

Particles that originate from the interaction vertex travel through the magnetic field and pass through all 
three regions of the drift chambers in a given sector and are reconstructed by tracking algorithms. First, 
in each super-layer adjacent wires with a signal are grouped into segments. Track candidates are constructed 
by connecting segments in each super-layer to form a track trajectory. Then each track candidate is fitted through 
a magnetic field to calculate the quality of the track ($\chi^2$), and the track candidates with $\chi^2$ below the cut value
are saved for further refinement using Kalman-filter~\cite{Kalman1960}.

The positions of these segments in each super-layer are used to fit the track trajectory 
to derive initial parameters, such as momentum and direction (called "hit-based" tracking).  After the initial selection, good track candidates 
(shown in Figure~\ref{fig:drift_chambers} with lines) are passed through a Kalman-filter to 
refine measured parameters further (called "time-based" tracking).

At the first stage of the tracking, the tracks are identified using only the hit positions (wire number). The precision 
of the reconstructed track parameters using only hit positions are worse ($\sim3\%$) than the final calculated parameters from time-based tracking, but they are sufficient to calculate physics quantities from the event and identify desired reactions in the event. 


The CLAS12 track reconstruction process is already making use of AI at different stages.
First, a Convolutional
Autoencoder is used to de-noise the drift chamber hits~\cite{Thomadakis:2022zcd}, which leads to improved cluster
identification in each super-layer. Then after the clustering, the track candidates are identified using a Multilayer perceptron (MLP) classifier,
which identifies 6-super-layer and 5-super-layer track candidates~\cite{Gavalian:2020oxg}. The combined result of
de-noising and AI-assisted track candidate identification used in CLAS12 leads to an increase in statistics of multi-particle 
final states of $50\%-75\%$ depending on kinematics~\cite{Gavalian:2020xmc}. 
Furthermore, an Artificial Intelligence approach to track reconstruction was developed to estimate track parameters 
(such as momentum and direction) based on cluster positions of the track~\cite{Thomadakis:2023ebe}, it was shown that 
the AI-estimated track parameters are closer to the values reconstructed using the Kalman-filter than the 
conventional hit-based tracking.

The developed AI tools for CLAS12 reconstruction can identify tracks from their cluster pattern and estimate the track parameters.
If combined with the clustering algorithm, this chain may provide a complete track reconstruction solely based on AI algorithms and
may potentially be used in real-time with data acquisition. 

\section{Track finding using Neural Networks approach}

\begin{figure}[h!]
\centering
\includegraphics[width=0.32\columnwidth]{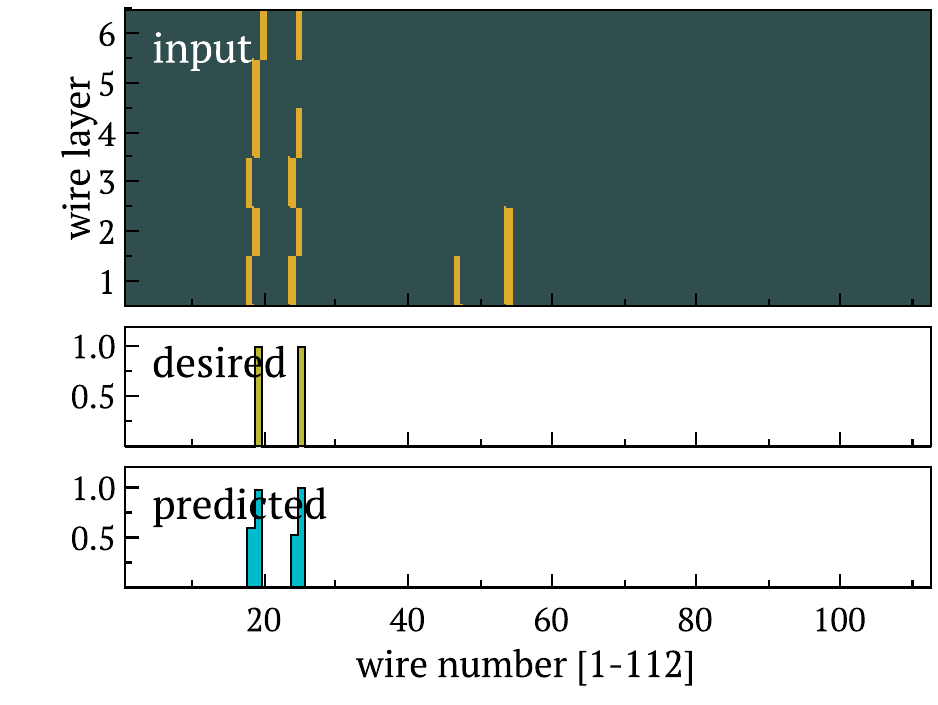}
\includegraphics[width=0.32\columnwidth]{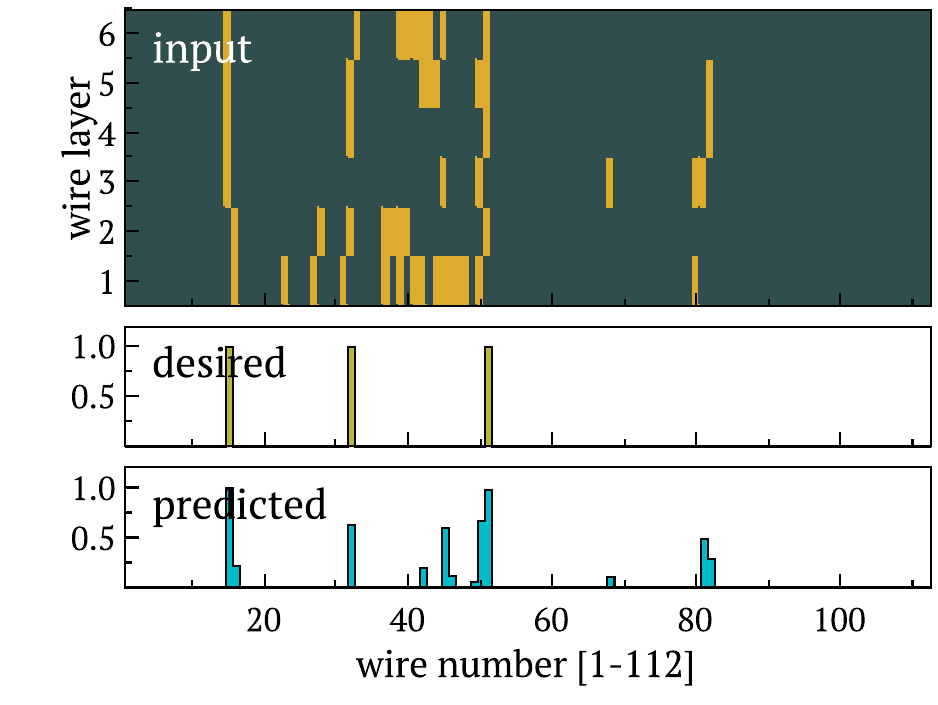}
\includegraphics[width=0.32\columnwidth]{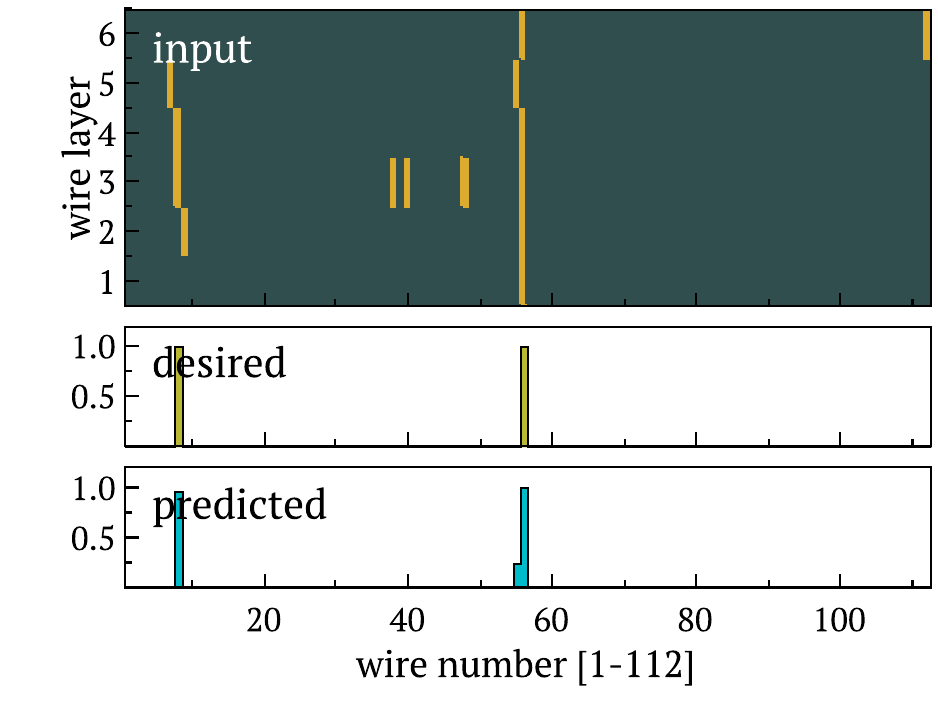}
\includegraphics[width=0.32\columnwidth]{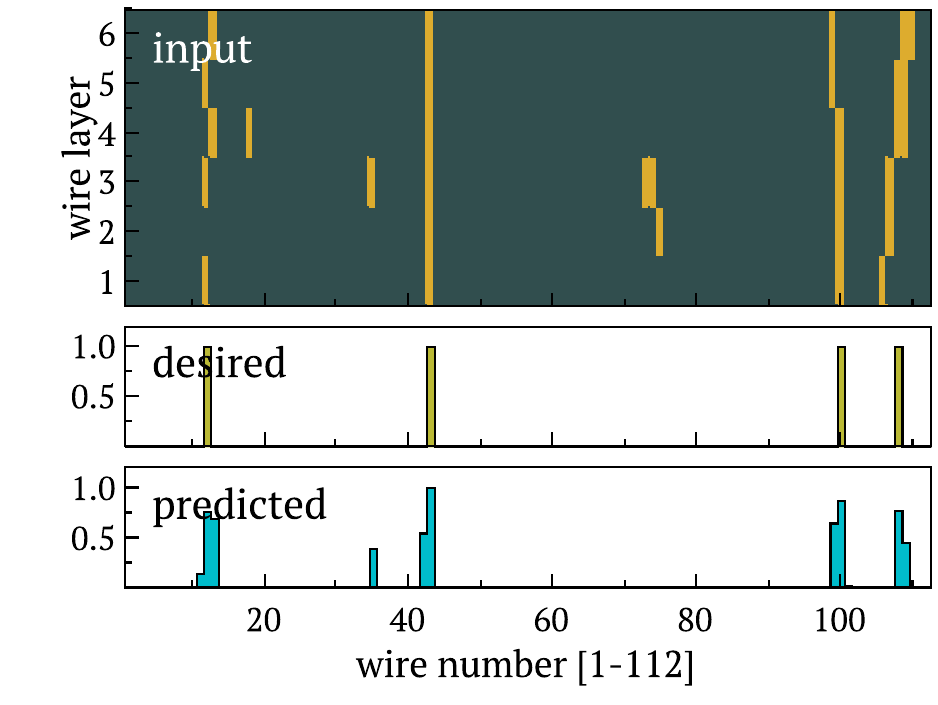}
\includegraphics[width=0.32\columnwidth]{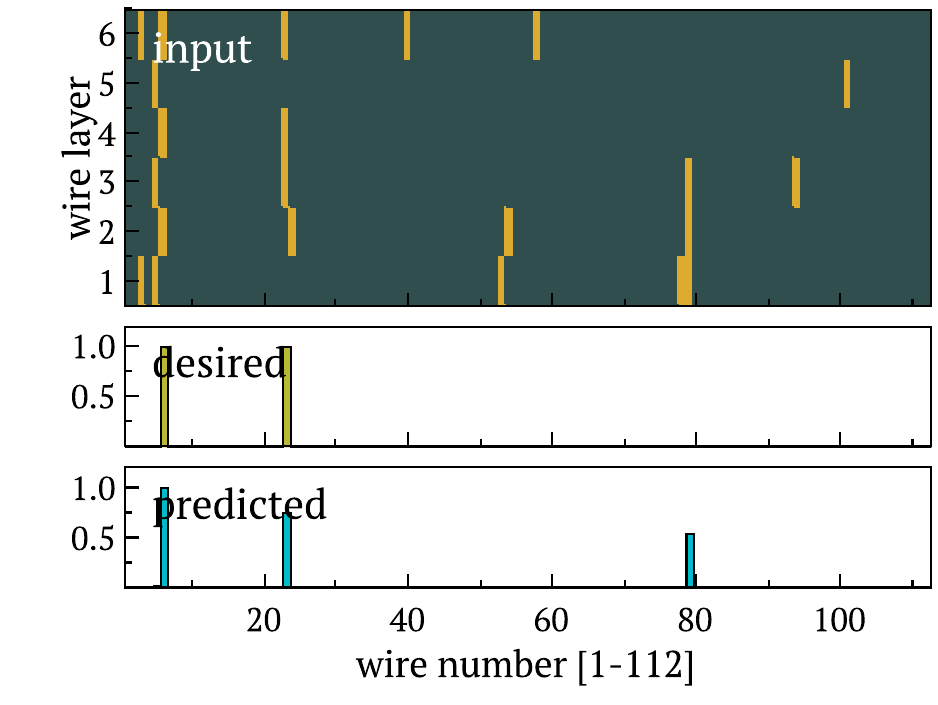}
\includegraphics[width=0.32\columnwidth]{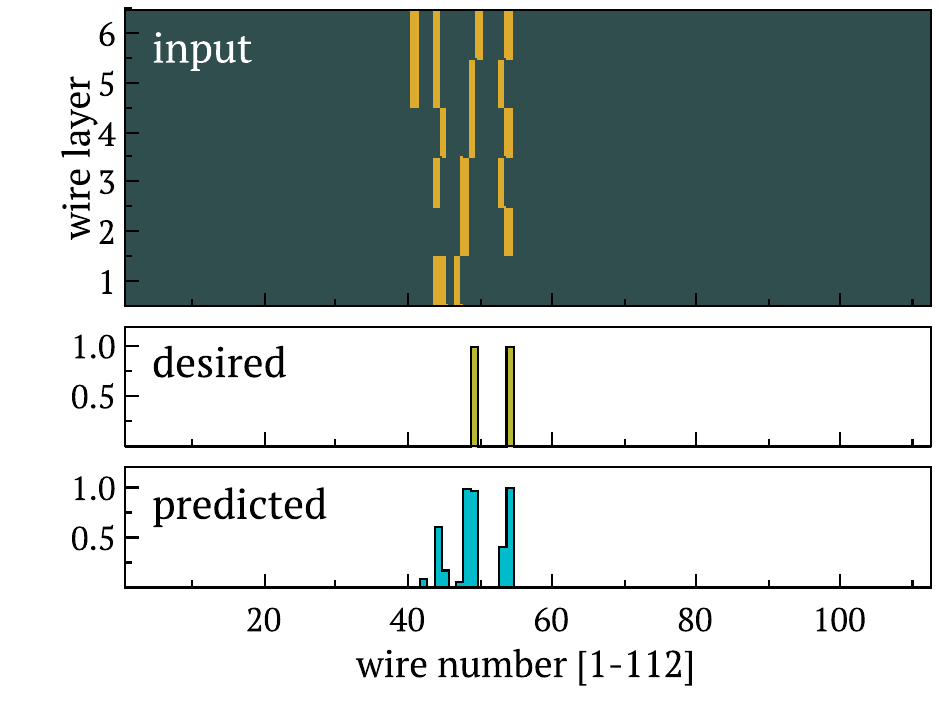}
\caption{Examples of cluster reconstructions using the CNN approach, the top plot (marked as "input") shows the input image of size $112\times6$ for the CNN, the histogram below (marked "desired") shows the reconstructed segments average wire position from the conventional segment finding algorithm, and the histogram below (marked  "predicted") shows the predicted segment position by the CNN.} 
\label{fig:segment_finding}
\end{figure}

Traditionally, the segments are identified in the tracking code by iterating over the hits in each super-layer, however, this algorithm requires the knowledge of the CLAS12 magnetic field and it is not very performant ($\sim 15 ms/event$). To achieve a real-time track reconstruction, a segment-finding neural network was developed using a Convolutional Logistic Regression Network.
The hit data from each drift chamber super-layer was presented as an image of size $112\times6$ used as an input to convolutional layers and the output was an array of $112$ numbers where the values of the bin corresponding to the position of the segment was set to 1.  The training sample was constructed using segments found by the conventional segment-finding algorithm. The input images and the output array can be seen in Figure~\ref{fig:segment_finding}. 

\begin{figure}[h!]
\centering
\includegraphics[width=0.23\columnwidth]{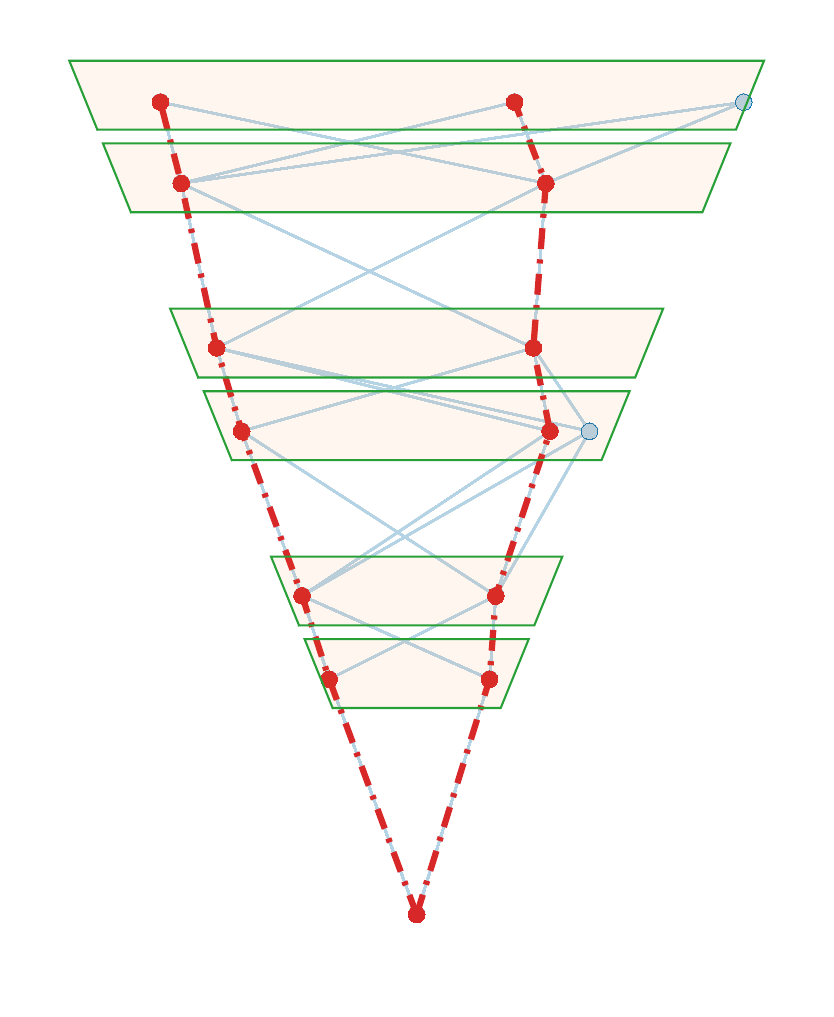}
\includegraphics[width=0.23\columnwidth]{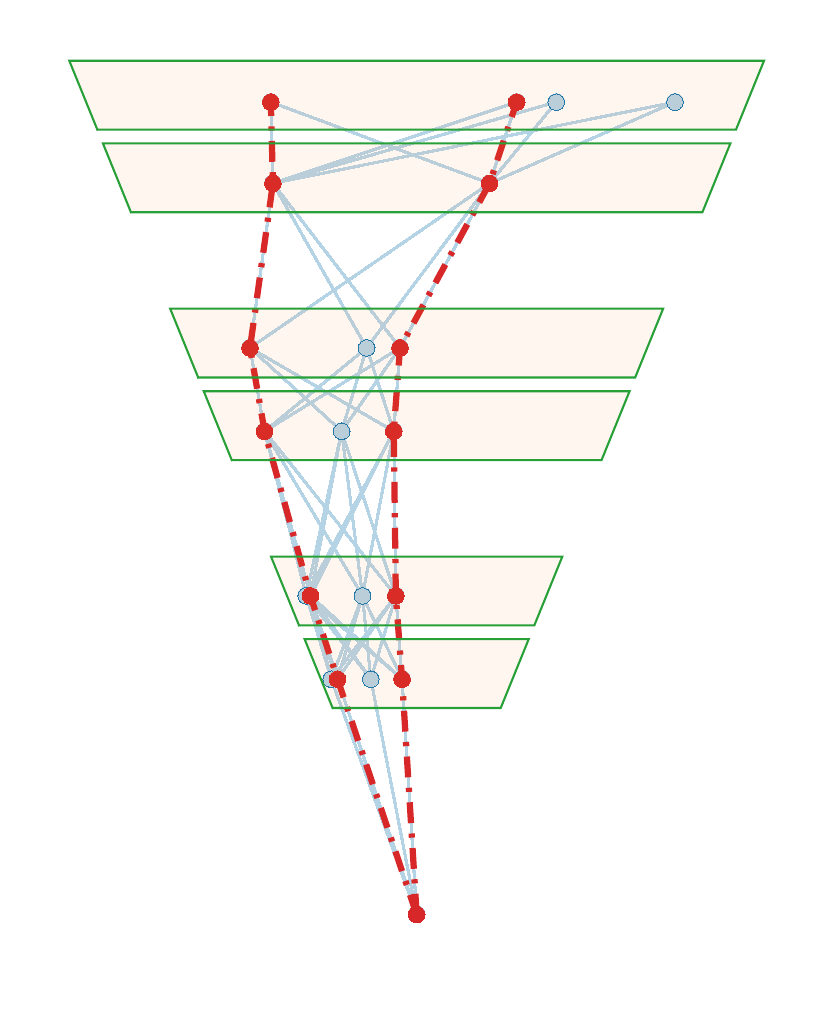}
\includegraphics[width=0.23\columnwidth]{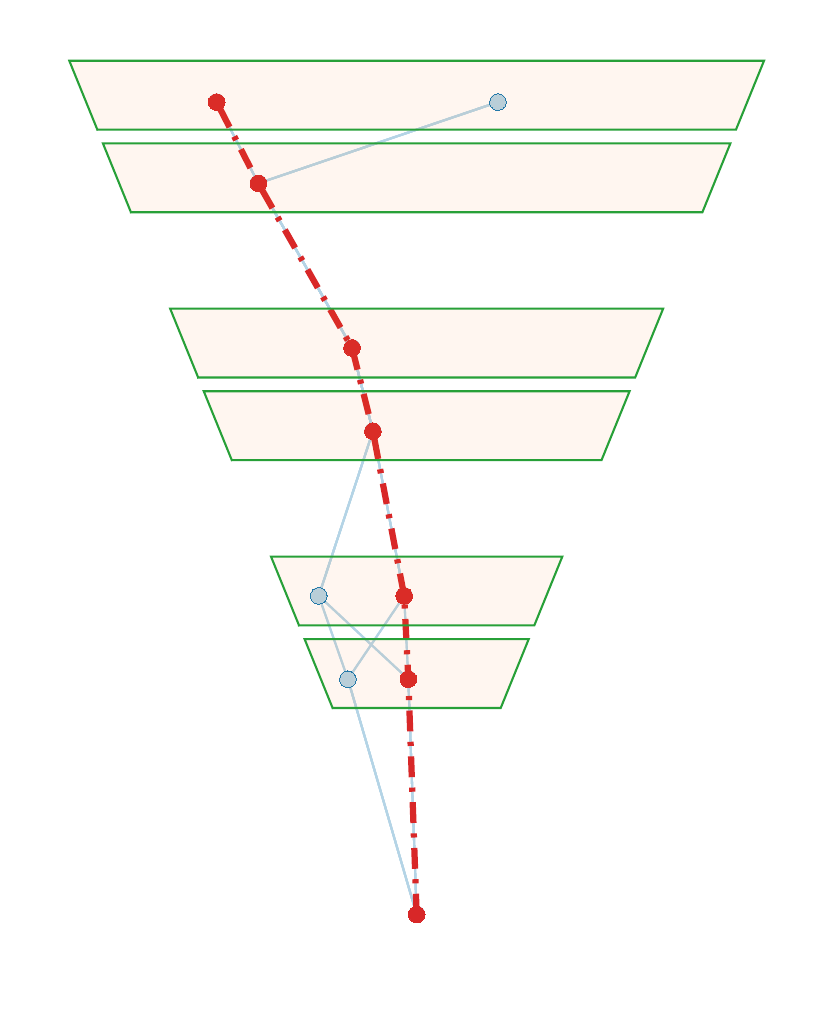}
\includegraphics[width=0.23\columnwidth]{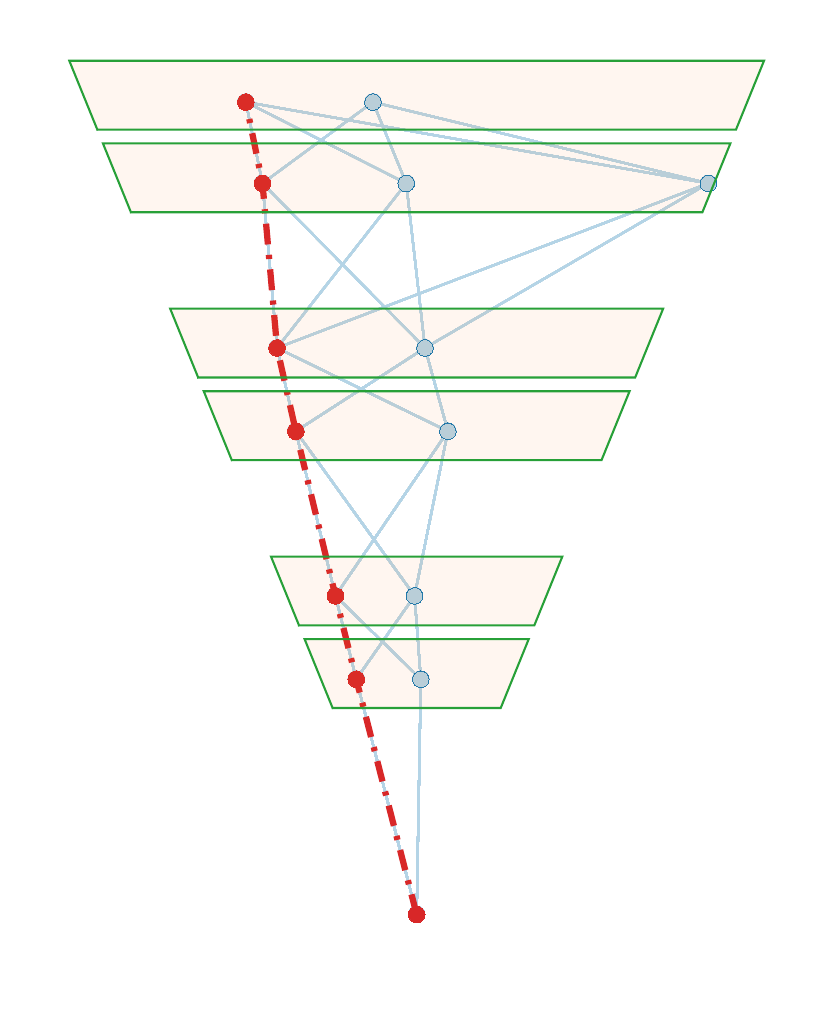}
\caption{Examples of track identification using MLP neural network. The solid lines connecting segments show all combinations of the tracks considered by the neural network and the dashed lines show tracks identified by the network.} 
\label{fig:track_classifier}
\end{figure}

After identifying clusters in each super-layer, the track classification neural network~\cite{Gavalian:2020oxg} is used to identify tracks in the event.
Examples of track classification can be seen in Figure~\ref{fig:track_classifier}. Once the tracks are identified in the event the track parameters (consisting of six numbers representing the segment positions in each super-layer) are passed to another neural network which predicts the particle momentum and direction~\cite{Thomadakis:2023ebe}.


By using three neural networks (a segment finder, a track classifier, and a tracker parameter estimator), the tracks in each event are reconstructed based only on hit positions in the drift chambers. 

\section{A reconstruction algorithm based on neural networks}

\begin{figure}[h!]
\centering
\includegraphics[width=0.85\columnwidth]{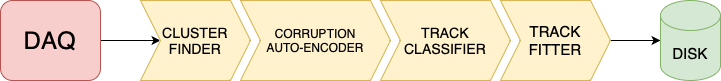}
\caption{The CLAS12 raw data is passed through four neural networks with the last one outputting the full parameters of the tracks (momentum and direction), which can be used for the identification of the event topology and  and physics analysis.} 
\label{fig:workflow}
\end{figure}

Combining the segment-finding neural network with our existing track-finding and parameter-extraction networks allowed for the construction of a workflow that is capable of fast-track reconstruction and can be used online. The schematic view of the workflow is shown in Figure~\ref{fig:workflow}.
The drift chamber data from the experimental setup was used to construct images ($112\times6$ arrays) filled with hits, and segments were identified 
using the segment-finding neural network. Then a track candidate list was constructed for each sector using segments from each of 6 super-layers.

\vspace{0.3in}

\begin{figure}[h!]
\centering
\includegraphics[width=0.95\columnwidth]{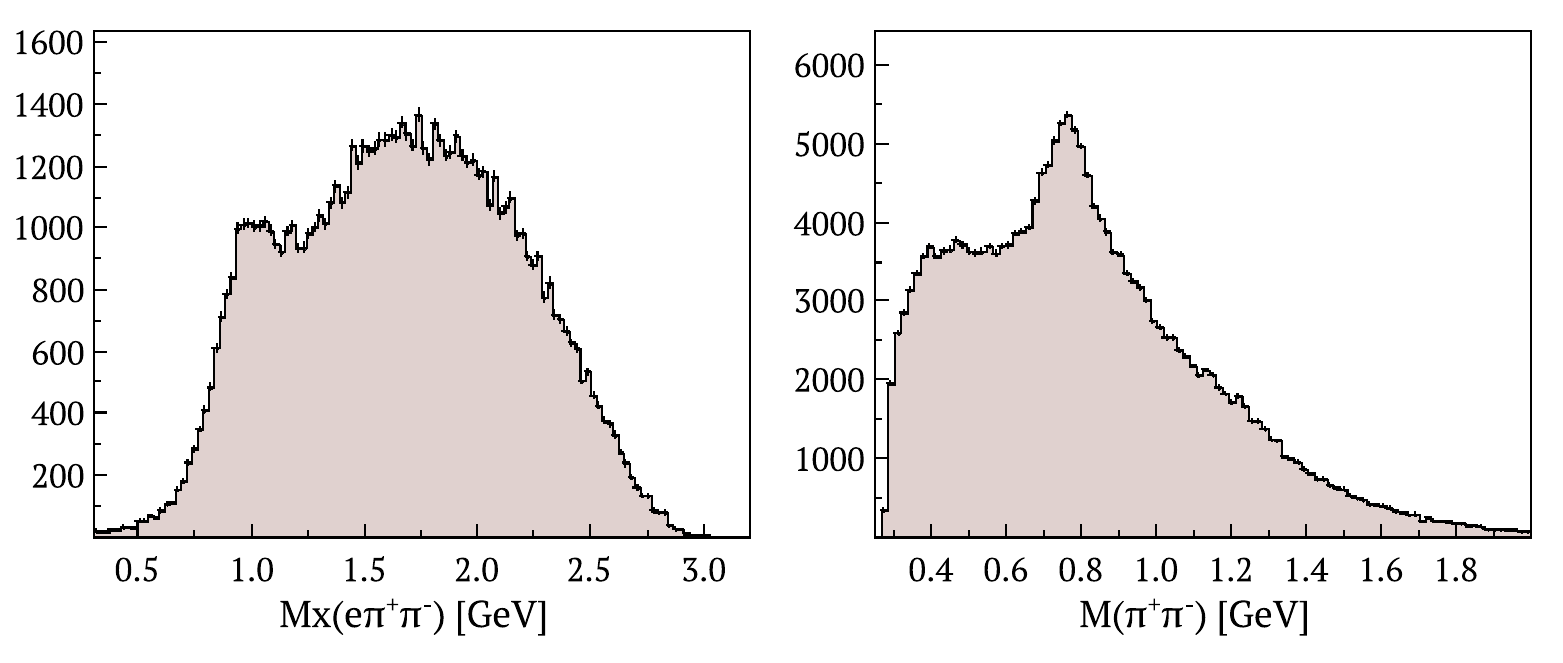}
\caption{The missing mass of $e(H)\rightarrow e^\prime\pi^+\pi^-X$ is shown with a proton peak visible at the mass value of $0.938~GeV$ (left). The invariant mass of $\pi^+ \pi^-$ is shown (right). The $\rho$ peak is visible at $770$ {\rm MeV}.  }
\label{fig:rho}
\end{figure}

The tracks were identified from the candidate list using track classification neural network~\cite{Thomadakis:2022zcd}, and then the track parameters were inferred using track parameter estimator neural network. In the output of the workflow, a data structure is created containing particles reconstructed in the event. The whole workflow reaches a speed of $450-500${~\rm Hz} in a single thread, using 48 cores we were able to run the reconstruction at approximately $22${~\rm kHz}, for comparison the experiment collects data at $12${~\rm kHz} (depending on the incident beam current and the target type).
Using reconstructed track parameters physics analysis can be performed to identify specific final states in the data. However, since no particle identification is performed at this level, some assumptions have to be made about the particle types when analyzing the data. 
In the following example, we isolated events that contain two negative and one positive tracks and extracted the signal for the $\rho$ meson decaying into a pion pair ($\rho\rightarrow \pi^+ \pi^-$). In one case it was assumed that the first negative particle is the electron and the second one is $\pi^-$ and in the second case the opposite. The positive track is assumed to be $\pi^+$. The invariant mass of two pions was constructed, the electron momentum was used to calculate the transferred momentum to $\rho^\circ$ and a cut was applied to clean the sample. The resulting distribution of the invariant mass of two pions (obtained by combining both combinatorial assumptions) can be seen in Figure~\ref{fig:rho} (right). 

Similarly, we can calculate the missing mass of the scattered electron and two pions to identify the missing proton, shown in Figure~\ref{fig:rho} (left). 
As it can be seen from the figure the $\rho$ meson is identified with good resolution ($\sim 2\%$) in the data stream without resorting to computationally extensive data processing procedures. The resolution of the extracted peak is worse than what is achieved by conventional methods after the time-based Kalman-filter tracking, but sufficient enough to isolate the events for further processing. 

\section{Discussion}

The developments discussed in this article have huge implications for the future of Nuclear Physics experiments, specifically for multi-purpose experiments, such as CLAS12 at JLab or ePIC at the future Electron Ion Collider. CLAS12 is a large acceptance spectrometer, and data collected during an experiment can be used to analyze (measure) many physics reactions. This presents some challenges in the procedure of obtaining the data. Traditionally, an entire dataset has to be processed by computer farms, which can typically take up to 6 months, and then the data has to be analyzed to extract the events of interest that match specific topology patterns (i.e. particles in the final state), which may comprise the fraction of the entire dataset. Additionally, when improvements are made to the reconstruction code that can improve the quality of the processed data, it is difficult to re-process the data set again due to limitations on allocated computing time. Many of the analyses end up analyzing data that is significantly less than a percent of the entire dataset. If the experimental data was tagged with the types of physics reactions in the events, the extraction and processing of specific data samples would be significantly faster and even possible to do on one computer over a weekend. (The estimate is based on our tests with an M3 Macbook Pro, capable of processing one run from an experiment in 20 minutes utilizing 10 threads, one could process 144 runs in one weekend, more than collected during one experiment). This can significantly decrease the computing requirements of an experiment and lead to massive energy savings.
A hierarchical data format, specifically designed to store events categorized by any key criteria, is used (HIPO) to store CLAS12 data~\cite{Gavalian:hipo}.
The online reconstruction will use the event tagging capability of HIPO to arrange events in the output stream according to their topology. This allows reading only relevant events to pass through reconstruction algorithms, minimizing the processing time of relevant events matching specific analysis criteria.
In conclusion, in the modern era of Artificial Intelligence and increased data volumes produced by Nuclear Physics experiments, similar approaches to identifying physics in the data stream and facilitating data processing and analysis workflows must be employed. This work is a step towards solving 
the challenges of optimal event identification and data storage for modern Nuclear Physics experiments.

\acknowledgments

This material is based upon work supported by the U.S. Department of Energy, Office of Science, Office of Nuclear Physics under contract DE-AC05-06OR23177, and
 NSF grant no. CCF-1439079 and the Richard T. Cheng Endowment. This work was performed using the Turing and  Wahab computing clusters at Old Dominion University.


\end{document}